Title

# Development of pericardial fat count images using a combination of three different deep-learning models

Authors

Takaaki Matsunaga, MD[a], Atsushi Kono, MD, PhD[a], Hidetoshi Matsuo, MD, PhD[a], Kaoru Kitagawa[b], Mizuho Nishio, MD, PhD[a,*], Hiromi Hashimura, MD, PhD[a], Yu Izawa, MD, PhD[c], Takayoshi Toba, MD, PhD[c], Kazuki Ishikawa[b], Akie Katsuki, BEng[d], Kazuyuki Ohmura, MEng[d], Takamichi Murakami, MD, PhD[a]

[a] Department of Radiology, Kobe University Graduate School of Medicine, Kobe, Japan

[b] Department of Radiation Technology, Kobe University Hospital, Kobe, Japan

[c] Division of Cardiovascular Medicine, Department of Internal Medicine, Kobe University Graduate School of Medicine, Kobe, Japan

[d] GE Healthcare Japan, Tokyo, Japan

*Corresponding author:

Mizuho Nishio, MD, PhD

Department of Radiology

Kobe University Graduate School of Medicine

7-5-2 Kusunoki-cho, Chuo-ku, Kobe 650-0017, Japan

Tel.: +81-78-382-6104

Fax: +81-78-382-6129

E-mail: nishiomizuho@gmail.com

**Abstract**

**Rationale and Objectives:** Pericardial fat (PF), the thoracic visceral fat surrounding the heart, promotes the development of coronary artery disease by inducing inflammation of the coronary arteries. For evaluating PF, this study aimed to generate pericardial fat count images (PFCIs) from chest radiographs (CXRs) using a dedicated deep-learning model.

**Materials and Methods:** The data of 269 consecutive patients who underwent coronary computed tomography (CT) were reviewed. Patients with metal implants, pleural effusion, history of thoracic surgery, or that of malignancy were excluded. Thus, the data of 191 patients were used. PFCIs were generated from the projection of three-dimensional CT images, where fat accumulation was represented by a high pixel value. Three different deep-learning models, including CycleGAN, were combined in the proposed method to generate PFCIs from CXRs. A single CycleGAN-based model was used to generate PFCIs from CXRs for comparison with the proposed method. To evaluate the image quality of the generated PFCIs, structural similarity index measure (SSIM), mean squared error (MSE), and mean absolute error (MAE) of (i) the PFCI generated using the proposed method and (ii) the PFCI generated using the single model were compared.

**Results:** The mean SSIM, MSE, and MAE were as follows: 0.856, 0.0128, and 0.0357, respectively, for the proposed model; and 0.762, 0.0198, and 0.0504, respectively, for the single CycleGAN-based model.

**Conclusion:** PFCIs generated from CXRs with the proposed model showed better performance than those with the single model. PFCI evaluation without CT may be possible with the proposed method.

**List of Abbreviations**

- 2D, two-dimensional
- 3D, three-dimensional
- AI, Artificial intelligence
- CT, computed tomography

- CWRS, CT-weighted ray sum image

- CXR, chest radiograph

- EAT, epicardial adipose tissue

- HU, Hounsfield Unit

- MAE, mean absolute error

- MSE, mean squared error

- PF, pericardial fat

- PFCI, pericardial fat count image

- SSIM, structural similarity index measure

**Introduction**

Epicardial adipose tissue (EAT) plays a crucial role in cardiac metabolism, mechanical protection of coronaries, and innervation [1,2]. EAT secretes anti-inflammatory cytokines that have protective effects on the coronary arteries and myocardium. However, these anti-inflammatory cytokines exert adverse lipotoxic effects under pathological conditions and is considered to be a cardiovascular risk factor. For instance, the secretion of inflammatory cytokines from EAT can directly affect the myocardium and cause adverse cardiovascular effects [3]. Previous studies have reported that an increase in EAT is associated with abnormalities in ventricular structure and function, coronary artery disease, and atrial fibrillation [4]. Similarly, pericardial fat (PF), which is defined as the thoracic visceral fat surrounding the heart, may also adversely affect the cardiovascular system.

Computed tomography (CT) is commonly used for the evaluation of PF. The evaluation of PF from CT images requires segmentation of PF, which may be performed manually or using dedicated software. However, radiation exposure during CT examinations is associated with adverse effects [5]. Therefore, performing CT solely for the evaluation of PF is discouraged due to ethical considerations. Minimally invasive and low-cost evaluation of PF in a large number of patients may contribute to a reduction in the incidence of cardiovascular events. Chest radiographs (CXRs), which are used commonly for mass screening, can be acquired at a low cost with low radiation exposure. However, since CXRs are two-dimensional (2D) images, the estimation of PF using these images is extremely difficult.

Artificial intelligence (AI) has been applied to medical image processing in recent years, and it has demonstrated good performance in various fields [6–15]. Previous studies have reported segmentation or evaluation of PF/EAT on CT images with and without the aid of AI [9,16–19]. However, to the best of our knowledge, no previous study has used CXRs for the evaluation of PF/EAT. Estimation of PF from CXRs using AI would be very useful. However, as CXRs are 2D images, the accurate evaluation of PF using a conventional AI model alone would be difficult. Therefore, for evaluating PF, this study aimed to determine whether our proposed method, which combines three AI models, can generate PF-related images with higher image quality than those of the conventional model.

**Materials and methods**

This single-center, retrospective study was approved by the Institutional Review Board of our hospital. The requirement for obtaining informed consent from the participants was waived owing to the retrospective nature of the study.

*Patients*

The data of 269 consecutive patients who underwent coronary CT between April 2020 and August 2021 were reviewed. Among these patients, 78 patients with metal implants (e.g., stent-grafts or pacemakers), pleural effusion, history of thoracic surgery, or that of malignancy were excluded. Thus, the data of 191 patients were included in this study. CXRs were acquired within a month of coronary CT. As CXRs were unavailable for some patients, a total of 191 CT images and 110 CXRs were collected.

*Dataset*

CXRs were acquired using a system (Digital Diagnost, Philips Medical Systems, the Netherlands). The cardiac CT examination was performed using a 256-detector row CT scanner (Revolution CT, GE HealthCare, US). Non-enhanced thoracic CT images were used for the evaluation. The CT image had a slice thickness of 1.25 mm and a resolution of 512 × 512. The tube voltage for the CT examination was 120 kV, and the tube current was controlled using an automated exposure control system.

PF was segmented by manually tracing the cardiac circumference on the CT images using the workstation (Ziostation2, Ziosoft, Japan). The voxels with CT values between -190 and -30 Hounsfield Unit (HU) were extracted as PF from the traced areas. A 2-cm cephalad slice of the left coronary artery origin, the diaphragm, and the anterior circumference of the esophagus were used as the upper, lower, and posterior borders of PF, respectively. Segmentation was initially performed by a radiation technologist with 7 years of experience. Subsequently, the segmentation results were confirmed and adjusted by a radiologist with 5 years of experience in clinical radiology.

*Proposed method*

We proposed the way of generating pericardial fat count images (PFCIs) to evaluate PF in this study. PFCI is a 2D image with the same slice plane as the coronal CT image. The number of voxels of PF was counted in the AP direction on the three-dimensional (3D) CT image to obtain PFCI from CT images, similar to the ray-sum image [20,21]. The PF count was used as the pixel value of PFCI. As a result, PFCI had the same slice plane as the coronal CT image. The ground truth of the PFCI was calculated via the manual segmentation of PF, as described in the subsection *Dataset*. Figure 1 presents sample images of CT and PFCI. The proposed method was developed to generate PFCIs from CXRs by combining three AI models. Figure 2 presents a schematic illustration of the proposed method for generating PFCIs from CXRs.

As the proposed method requires pericardial segmentation, a 2D projection of the PF tracing result in the coronal plane was used as the ground truth for pericardial segmentation. Furthermore, the CT values of the CT images were used to compute the projected image of the radiographic attenuation in the coronal plane in the proposed method. The projected images were referred to as CT-weighted ray sum images (CWRSs). CXRs were converted to CWRSs using our deep-learning model. This enabled other deep-learning models to perform prediction or image translation in the CT domain rather than directly in the CXR domain.

The steps of the proposed method to generate PFCIs from CXRs are as follows.

(1) CWRS was generated from CXR.

(2) Pericardial segmentation was performed on the generated CWRS to generate the pericardial mask.

(3) PFCI was generated from CWRS with the pericardial mask.

The following three models were developed in the proposed method for steps (1)–(3):

(I) CycleGAN [15] was used to convert CXRs into CWRSs.

(II) The pericardial region was segmented from CWRSs using U-Net [14].

(III) Pix2Pix [11] was used to generate PFCIs from CWRSs with the pericardial mask.

The models required the following datasets: (i) unpaired CWRS and CXR, (ii) paired CWRS and pericardial segmentation results, and (iii) paired CWRS with pericardial mask and PFCI. The results of the proposed method were compared with those of a CycleGAN-based single model that directly generated PFCIs from CXRs. The details of the models and image processing are described below. Nested cross-validation was used in the proposed method to prevent data leakage and overfitting.

*Generation of CWRSs from CXRs*

This subsection describes the model built for the generation of CWRSs from CXRs. As a preprocessing step for the CT images, CWRSs were generated from the 3D CT images by adding the CT values (HU) multiplied by an attenuation factor in the AP direction for each voxel. The following equation was used to generate CWRS:

$$I_{xz} = \sum_{y=1}^{N} CT_{xyz}(1 - 0.0018y)/N$$

(1),

where $CT_{xyz}$ represents the CT value (HU) of the 3D CT images at the coordinates $x, y, z$; $I_{xz}$ represents the pixel value of CWRS at $x, z$; and $N$ represents the length of the 3D CT images in the $y$ direction ($N$ = 512). Voxels below -1000 HU (e.g., air) were excluded from the calculation when generating CWRSs. The attenuation coefficient $(1 - 0.0018y)$ was empirically determined such that the attenuation coefficient was not negative at the last voxel (512$^{th}$ voxel), and the image quality of CWRS was close to that of CXR.

CycleGAN was employed to generate CWRSs from CXRs (Domain A, CXR; Dmain B,CWRS). Compared with that of the original CycleGAN model, the number of channels in the input and output images was changed to 1 in the CycleGAN model in this study. The four optimizers for the two sets of generators and discriminators were Adam, and their learning rates were set to 0.0002. The number of training epochs was 200. The batch size was 1. Min/max normalization was applied to domains A and B such that the range of the input image was between -1 and 1. In addition, the input and output images were resized to 256 × 256 pixels.

*Pericardial segmentation in CWRSs*

This subsection describes the model built for the pericardial segmentation of CWRSs. The input image was a 256 × 256 CWRS, and the ground truth of the pericardial segmentation was a 256 × 256 2D mask created from the PF tracing results of the CT image. This input/ground-truth pair was used to train the 2D U-Net. U-Net has eight connections between the contracting and expanding paths. The loss function of the U-Net was the sum of the binary cross-entropy and dice loss. The optimizer used was Adam, with a learning rate of 0.001. The learning rate was set to 0.001. The number of training epochs and the batch size were set to 100 and 12, respectively.

*Generation of PFCI from CWRSs with pericardial mask*

This subsection describes the model built for the generation of PFCIs. The pericardial mask obtained from the segmentation described in the subsection *Pericardial segmentation in CWRSs* was used in this model. A CWRS with a pericardial mask was obtained. The ground truth of the PFCI was calculated from the results of the manual segmentation of PF on a 3D CT image. Image translation from CWRS with the pericardial mask to PFCI was performed using Pix2Pix (Domain A, CWRS with the pericardial mask; Domain B, PFCI). The optimizer of Pix2pix was Adam, with a learning rate of 0.001. The loss function of Pix2pix was the weighted sum of the conditional GAN loss and L1 loss with weights of 1 and 10, respectively. For image transformation from Domains A to B, Pix2pix was trained with a patch size of PatchGAN 96, a batch size of 10, and 400 training epochs.

*Control model*

PFCIs were directly generated from CXRs using a single CycleGAN model (Domain A: CXR, Domain B: PFCI). The details of the single-CycleGAN model are the same as those described in the subsection *Generation of CWRSs from CXRs*. Figure 3 presents a schematic illustration of the single CycleGAN model used to generate PFCIs from CXRs. Ten-fold cross-validation was performed to train and evaluate the control model.

**Evaluation**

For the 110 cases with both CT images and CXRs, PFCIs were generated from CXRs using nested cross-validation for the proposed method and cross-validation for the control model. This resulted in the following three PFCIs for each CXR: the ground truth of PFCI from CT, the PFCI generated using the proposed method, and the PFCI generated using the control model. Using these three PFCIs, the mean absolute error (MAE) and mean squared error (MSE) were calculated based on the following equations:

$$MAE = \sum_{x=1}^{N}\sum_{z=1}^{M} \frac{|PFCI_{xz}^{GT} - PFCI_{xz}^{M}|}{MN}$$

(2)

$$MSE = \sum_{x=1}^{N}\sum_{z=1}^{M} \frac{|PFCI_{xz}^{GT} - PFCI_{xz}^{M}|^2}{MN}$$

(3)

where $PFCI_{xz}^{GT}$ represents the pixel value of the ground truth of the PFCI at the coordinates *x, z*; and $PFCI_{xz}^{M}$ represents the pixel value of the generated PFCI at coordinates *x,z*. MAE and MSE were calculated between (i) the PFCI generated using the proposed method and the ground truth and (ii) the PFCI generated using the control model and the ground truth. The smaller the MAE and MSE values, the closer the generated PFCI was to the ground truth. In addition, the structural similarity index

measure（SSIM）[22] was calculated between the model output and ground truth using the following equation, and the generated PFCI with a higher SSIM was considered closer to the ground truth:

$$SSIM(x,y) = \frac{(2u_x u_y + c_1)(2s_{xy} + c_2)}{(u_x^2 + u_y^2 + c_1)(s_{x2} + s_{y2} + c_2)}$$

(4)

where $x$ and $y$ are the generated PFCI and its ground truth, respectively; $u_x$ and $u_y$ are the means of $x$ and $y$, respectively; $s_{x2}$ and $s_{y2}$ are the variances of $x$ and $y$, respectively; $s_{xy}$ is the covariance of $x$ and $y$; and $c_1$ and $c_2$ are determined by the dynamic range of the pixel values to stabilize the division with the weak denominator. Before calculating SSIM, MSE, and MAE, the three PFCIs were normalized such that the pixel value was within the range of 0–1.

**Results**

Table 1 presents the characteristics of the 191 patients included in this study. Among the 191 patients included, 124 were males, and 67 were females. The mean age of the patients was 72.0 (interquartile range: 62.0–78.0) years. Both CT images and CXRs were available for 110, whereas only CT images were available for the remaining 81. The mean interval between the acquisition of the CT images and CXRs was 7.0 (3.0–14.0) days for the 110 patients.

Table 2 presents the SSIM, MSE, and MAE for the proposed and control models. The mean and standard deviation of SSIM, MSE, and MAE were as follows: 0.856±0.0244, 0.0128±0.00362, and 0.0357±0.00747, respectively, for the proposed method, and 0.762±0.0601, 0.0198±0.0124, and 0.0504±0.0174, respectively, for the control model. These quantitative results reveal that the image quality of the PFCIs generated using the proposed method was better than that of those generated using the control model.

Figures 4–7 present representative examples of CXRs, the ground truths of the PFCIs, and the PFCIs generated using the proposed and control models. These figures show that the image quality of the PFCIs generated using the proposed method were better than that of those generated using the control model. The distribution and overall shape of PF were distorted in the PFCIs generated using the single CycleGAN model.

**Discussion**

We developed deep-learning models to generate PFCIs from CXRs in this study. For this purpose, the proposed method combined three deep-learning models. The SSIM, MSE, and MAE of the PFCI generated using the control model (single CycleGAN model) and those of the PFCI generated using the proposed method were compared. The SSIM, MSE, and MAE of the PFCI generated using the proposed method were found to be superior to those of the PFCI generated using the control model. This indicates that our method can generate a PFCI closer to the ground truth of PFCI.

Compared with the single CycleGAN model, the image quality of PFCIs generated using the proposed method, which combined the three models, was better. It is difficult to match the shapes between the two domains for a single image using CycleGAN [10]. Consequently, the distribution and overall shape of the PF were distorted in the PFCI generated using the single-CycleGAN model. The proposed method addresses this problem by performing pericardial segmentation.

Although several previous studies have used one or two deep-learning models, studies that combined more than three models are relatively rare. As it is difficult to effectively combine three or more models, the fact that three models were combined in this study is a major strength.

CWRSs were utilized in this study. CWRSs are potentially useful in image preprocessing for deep learning using chest CT images and CXRs. Previous studies have generated CXRs from CT images or CT images from CXRs [13,23]; however, the novelty of this study lies in that we mapped both the image and its annotation results from the CT domain to the CXR domain.

This study has several limitations. First, we did not evaluate or predict the prognosis or cardiovascular events using the PFCI results. Second, we did not externally validate the proposed method.

**Conclusion**

In this study, PFCIs were generated from CXRs using the proposed method. CXRs are widely used in clinical settings. The use of the method developed in this study may enable PFCI evaluation in patients without the use of CT. The PFCIs generated using our proposed method could also aid in the prediction of cardiovascular events.

# References

1.  Wu Y, Zhang A, Hamilton DJ, Deng T. Epicardial Fat in the Maintenance of Cardiovascular Health. Methodist Debakey Cardiovasc J. 2017;13: 20–24. doi:10.14797/MDCJ-13-1-20

2.  Braescu L, Gaspar M, Buriman D, Aburel OM, Merce AP, Bratosin F, et al. The Role and Implications of Epicardial Fat in Coronary Atherosclerotic Disease. J Clin Med. 2022;11: 4718. doi:10.3390/JCM11164718

3.  Sacks HS, Fain JN. Human epicardial adipose tissue: A review. Am Heart J. 2007;153: 907–917. doi:10.1016/J.AHJ.2007.03.019

4.  Oba K, Maeda M, Maimaituxun G, Yamaguchi S, Arasaki O, Fukuda D, et al. Effect of the Epicardial Adipose Tissue Volume on the Prevalence of Paroxysmal and Persistent Atrial Fibrillation. Circ J. 2018;82: 1778–1787. doi:10.1253/CIRCJ.CJ-18-0021

5.  Preston DL, Mattsson A, Holmberg E, Shore R, Hildreth NG, Boice JD. Radiation Effects on Breast Cancer Risk: A Pooled Analysis of Eight Cohorts. https://doi.org/101667/0033-7587(2002)158[0220:REOBCR]20CO;2. 2002;158: 220–235. doi:10.1667/0033-7587(2002)158

6.  Nishio M, Kobayashi D, Nishioka E, Matsuo H, Urase Y, Onoue K, et al. Deep learning model for the automatic classification of COVID-19 pneumonia, non-COVID-19 pneumonia, and the healthy: a multi-center retrospective study. Sci Rep. 2022;12: 8214. doi:10.1038/S41598-022-

11990-3

7. Matsuo H, Nishio M, Kanda T, Kojita Y, Kono AK, Hori M, et al. Diagnostic accuracy of deep-learning with anomaly detection for a small amount of imbalanced data: discriminating malignant parotid tumors in MRI. Sci Rep. 2020;10: 19388. doi:10.1038/S41598-020-76389-4

8. Rodrigues O, Pinheiro VHA, Liatsis P, Conci A. Machine learning in the prediction of cardiac epicardial and mediastinal fat volumes. Comput Biol Med. 2017;89: 520–529. doi:10.1016/J.COMPBIOMED.2017.02.010

9. Greco F, Salgado R, Van Hecke W, Del Buono R, Parizel PM, Mallio CA. Epicardial and pericardial fat analysis on CT images and artificial intelligence: a literature review. Quant Imaging Med Surg. 2022;12: 2075–2089. doi:10.21037/QIMS-21-945

10. Matsuo H, Nishio M, Nogami M, Zeng F, Kurimoto T, Kaushik S, et al. Unsupervised-learning-based method for chest MRI-CT transformation using structure constrained unsupervised generative attention networks. Sci Rep. 2022;12: 11090. doi:10.1038/S41598-022-14677-X

11. Isola P, Zhu JY, Zhou T, Efros AA. Image-to-image translation with conditional adversarial networks. Proceedings - 30th IEEE Conference on Computer Vision and Pattern Recognition, CVPR 2017. 2017;2017-January: 5967–5976. doi:10.1109/CVPR.2017.632

12. Moribata Y, Kurata Y, Nishio M, Kido A, Otani S, Himoto Y, et al. Automatic segmentation of bladder cancer on MRI using a convolutional neural network and reproducibility of radiomics

features: a two-center study. Sci Rep. 2023;13: 628. doi:10.1038/S41598-023-27883-Y

13. Ying X, Guo H, Ma K, Wu J, Weng Z, Zheng Y. X2CT-gan: Reconstructing CT from biplanar X-rays with generative adversarial networks. Proceedings of the IEEE Computer Society Conference on Computer Vision and Pattern Recognition. 2019;2019-June: 10611–10620. doi:10.1109/CVPR.2019.01087

14. Ronneberger O, Fischer P, Brox T. U-net: Convolutional networks for biomedical image segmentation. Lecture Notes in Computer Science (including subseries Lecture Notes in Artificial Intelligence and Lecture Notes in Bioinformatics). 2015;9351: 234–241. doi:10.1007/978-3-319-24574-4_28/COVER

15. Zhu JY, Park T, Isola P, Efros AA. Unpaired Image-to-Image Translation Using Cycle-Consistent Adversarial Networks. Proceedings of the IEEE International Conference on Computer Vision. 2017;2017-October: 2242–2251. doi:10.1109/ICCV.2017.244

16. de Albuquerque VHC, de A. Rodrigues D, Ivo RF, Peixoto SA, Han T, Wu W, et al. Fast fully automatic heart fat segmentation in computed tomography datasets. Computerized Medical Imaging and Graphics. 2020;80: 101674. doi:10.1016/J.COMPMEDIMAG.2019.101674

17. Rebelo AF, Ferreira AM, Fonseca JM. Automatic epicardial fat segmentation and volume quantification on non-contrast cardiac Computed Tomography. Computer Methods and Programs in Biomedicine Update. 2022;2: 100079. doi:10.1016/J.CMPBUP.2022.100079

18. Zhang Q, Zhou J, Zhang B, Jia W, Wu E. Automatic Epicardial Fat Segmentation and Quantification of CT Scans Using Dual U-Nets with a Morphological Processing Layer. IEEE Access. 2020;8: 128032–128041. doi:10.1109/ACCESS.2020.3008190

19. Benčević M, Galić I, Habijan M, Pižurica A. Recent Progress in Epicardial and Pericardial Adipose Tissue Segmentation and Quantification Based on Deep Learning: A Systematic Review. Applied Sciences 2022, Vol 12, Page 5217. 2022;12: 5217. doi:10.3390/APP12105217

20. Deb S, Lu Z, Kuganesan A, Lau KK. Ray sum image: its efficacy in renal tract calculus detection. Clin Radiol. 2019;74: 650.e7-650.e12. doi:10.1016/J.CRAD.2019.03.022

21. Seo H, Lee KH, Hyuk JK, Kim K, Kang SB, So YK, et al. Diagnosis of acute appendicitis with sliding slab ray-sum interpretation of low-dose unenhanced CT and standard-dose i.v. contrast-enhanced CT scans. AJR Am J Roentgenol. 2009;193: 96–105. doi:10.2214/AJR.08.1237

22. Wang Z, Bovik AC, Sheikh HR, Simoncelli EP. Image quality assessment: From error visibility to structural similarity. IEEE Transactions on Image Processing. 2004;13: 600–612. doi:10.1109/TIP.2003.819861

23. MATSUBARA N, TERAMOTO A, SAITO K, FUJITA H. Generation of Pseudo Chest X-ray Images from Computed Tomographic Images by Nonlinear Transformation and Bone Enhancement. Medical Imaging and Information Sciences. 2019;36: 141–146. doi:10.11318/MII.36.141

**Figure and Figure Legends**

Figure 1. Steps of image preprocessing for the pericardial fat count image.

Abbreviation: CT, computed tomography.

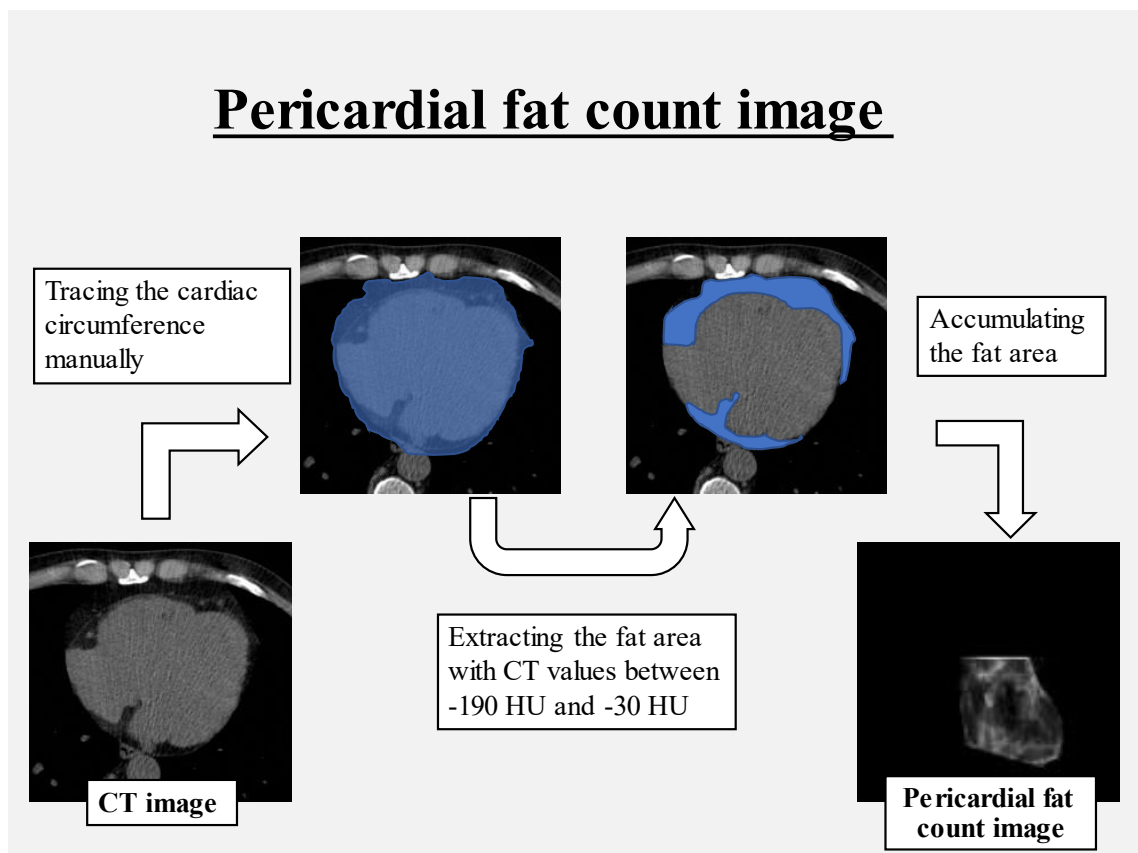

Figure 2. Combined model and the generated pericardial fat count image.

Abbreviation: CT, computed tomography.

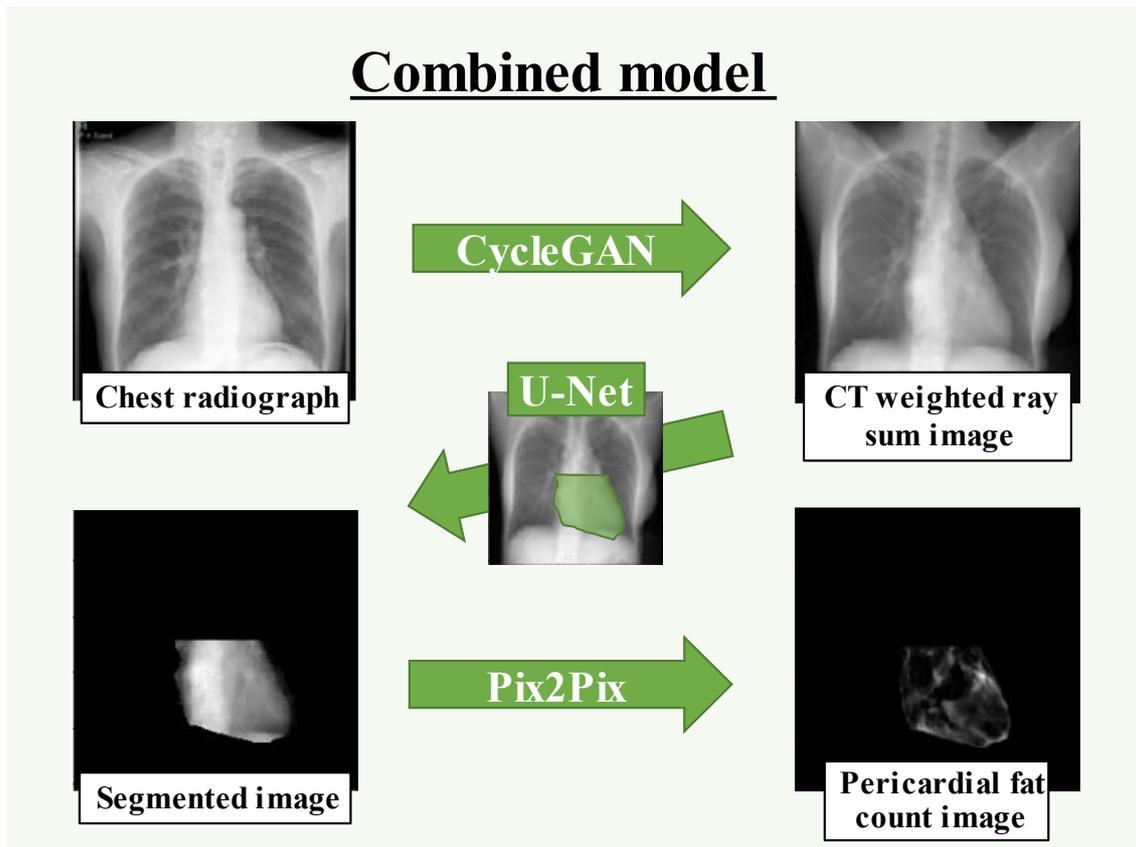

Figure 3. Control model and the generated pericardial fat count image.

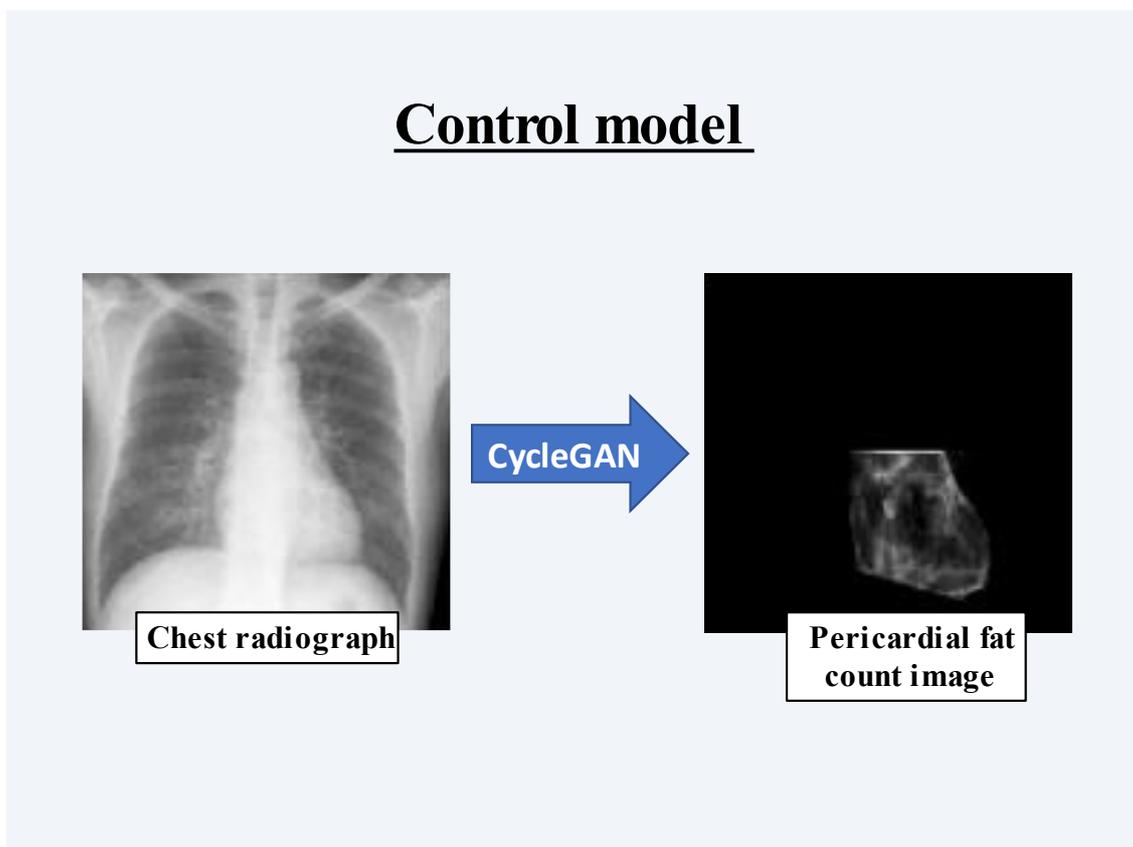

Figure 4. CXR, ground truth of PFCI, PFCI generated using the proposed method, and PFCI generated using the single model of a 65-year-old woman.

Abbreviation: CXR, chest radiograph; PFCI, pericardial fat count image

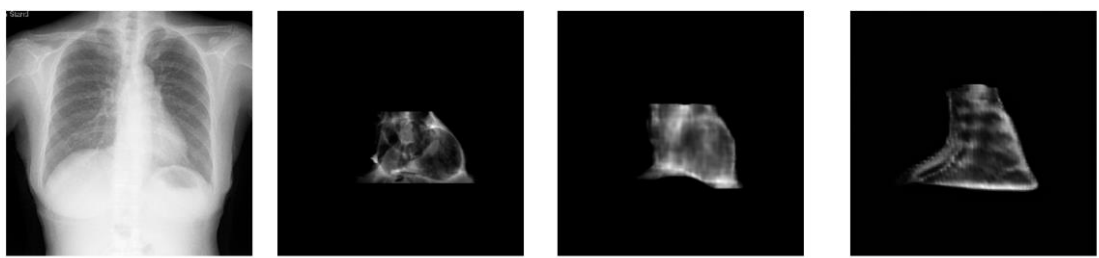

Figure 5. CXR, ground truth of PFCI, PFCI generated using the proposed method, and PFCI generated using the single model of a 64-year-old man.

Abbreviation: CXR, chest radiograph; PFCI, pericardial fat count image

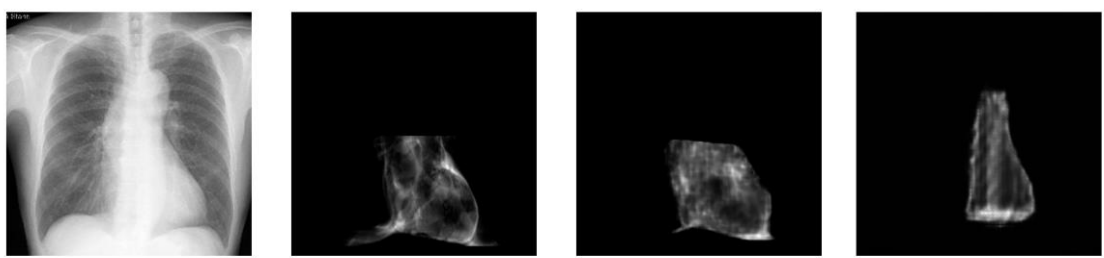

Figure 6. CXR, ground truth of PFCI, PFCI generated using the proposed method, and PFCI generated using the single model of a 75-year-old man.

Abbreviation: CXR, chest radiograph; PFCI, pericardial fat count image

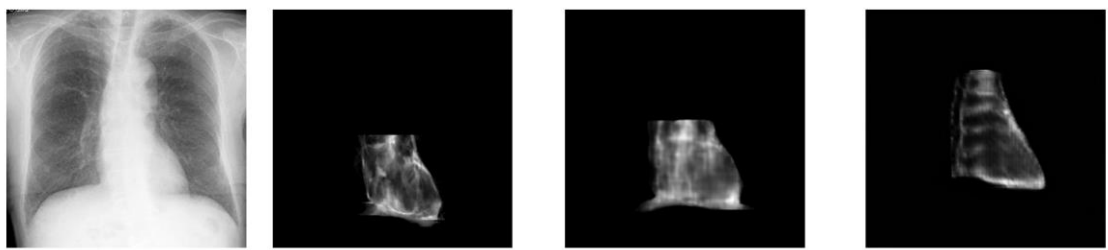

Figure 7. CXR, ground truth of PFCI, PFCI generated using the proposed method, and

PFCI generated using the single model of a 71-year-old woman.

Abbreviation: CXR, chest radiograph; PFCI, pericardial fat count image

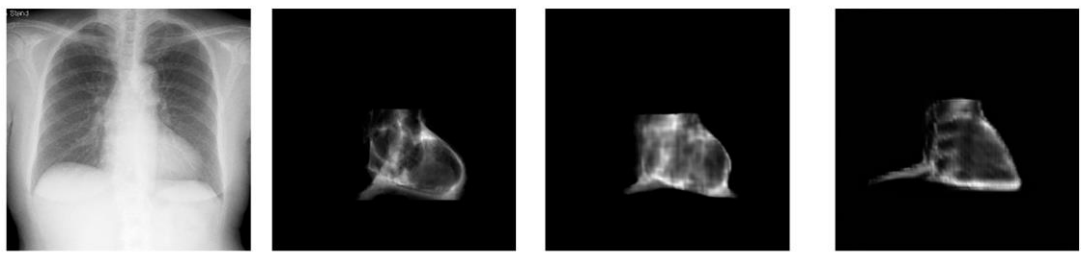

Tables

Table 1. Patient Characteristics

| Data | | Median or number of patients | (IQR or ratio) |
|---|---|---|---|
| **All** | | | |
| N | | 191 | |
| age | | 72 | (62.0-78.0) |
| gender | | | |
| | male | 124 | (64.9%) |
| | female | 67 | (35.1%) |
| **CT + CXR** | | | |
| N | | 110 | |
| age | | 73 | (64.5-79.0) |
| gender | | | |
| | male | 78 | (70.9%) |
| | female | 32 | (29.1%) |
| Interval between CT and CXR (day) | | 7.0 | (3.0-14.0) |
| **CT only** | | | |
| N | | 81 | |
| age | | 69 | (60.0-76.0) |
| gender | | | |
| | male | 46 | (56.8%) |
| | female | 35 | (43.2%) |

Abbreviation: interquartile range, IQR; computed tomography, CT; chest radiograph,

CXR.

Table 2. Results of quantitative metrics in the combined model and single model.

|      | Single model |        |        | Combined model |         |         |
|------|--------------|--------|--------|----------------|---------|---------|
|      | SSIM         | MSE    | MAE    | SSIM           | MSE     | MAE     |
| Mean | 0.762        | 0.0198 | 0.0504 | 0.856          | 0.0128  | 0.0357  |
| SD   | 0.0601       | 0.0124 | 0.0174 | 0.0244         | 0.00362 | 0.00747 |

Abbreviation: standard deviation, SD; mean absolute error, MAE; mean squared error, MSE; structural similarity index measure, SSIM.